\documentclass[aps,prx,reprint,longbibliography]{revtex4-2}

\usepackage{amsfonts,amscd,mathrsfs,amsmath,amsthm,amssymb}
\usepackage{mathtools}
\usepackage{xparse}
\usepackage{graphicx}
\usepackage{float}
\usepackage{pifont}
\usepackage[dvipsnames]{xcolor}
\usepackage{colortbl}
\usepackage{multirow}
\usepackage{enumerate}   
\usepackage{comment}
\usepackage{booktabs}
\usepackage{cancel}
\usepackage{lieart}

\usepackage{listings}
\usepackage{xcolor}

\definecolor{codegreen}{rgb}{0,0.6,0}
\definecolor{codegray}{rgb}{0.2,0.2,0.2}
\definecolor{codepurple}{rgb}{0.58,0,0.82}
\definecolor{backcolour}{rgb}{0.95,0.95,0.92}

\lstdefinestyle{mystyle}{
    backgroundcolor=\color{backcolour},   
    commentstyle=\color{codegreen},
    keywordstyle=[1]\bfseries\color{magenta},
    keywordstyle=[2]\color{blue},
    otherkeywords = {PerfectGroup,CharacterTable,Irr,Norm,ScalarProduct,gap>},
    morekeywords = [1]{PerfectGroup,CharacterTable,Irr,Norm,ScalarProduct,Degree},
    morekeywords = [2]{gap>,>},
    numberstyle=\tiny\color{codegray},
    stringstyle=\color{codepurple},
    basicstyle=\ttfamily\footnotesize,
    breakatwhitespace=false,         
    breaklines=true,                 
    captionpos=b,                    
    keepspaces=true,                 
    numbers=left,                    
    numbersep=5pt,   
    frame = single, 
    showspaces=false,                
    showstringspaces=false,
    showtabs=false,  
    morecomment=[l][\color{codegreen}]{\#},
    alsoletter=?>,
    tabsize=2
}
\PassOptionsToPackage{hyphens}{url}\usepackage{hyperref}

\hypersetup{
    colorlinks=true,
    linkcolor=magenta!80!black,  
    citecolor=magenta!80!black,
}
\usepackage{url} 

\usepackage[capitalise]{cleveref}
\usepackage{physics}
\usepackage{bm}
\usepackage{bbm}
\usepackage{caption}

\theoremstyle{theorem}

\theoremstyle{definition}

\newtheorem*{conjecture*}{Conjecture}

\AddToHook{cmd/appendix/before}{%
  \setcounter{equation}{0}%
  \renewcommand{\theequation}{\thesection\arabic{equation}}%
  \setcounter{theorem}{0}%
  \setcounter{lemma}{0}%
}

\newcommand\numberthis{\addtocounter{equation}{1}\tag{\theequation}}

    \DeclareMathAlphabet{\mathsfit}{T1}{\sfdefault}{\mddefault}{\sldefault}
    \SetMathAlphabet{\mathsfit}{bold}{T1}{\sfdefault}{\bfdefault}{\sldefault}

    \renewcommand{\S}{\mathsf{S}}

    \renewcommand{\SU}{\mathrm{SU}}
        \newcommand{\PSU}{\mathrm{PSU}}
\renewcommand{\Sp}{\mathrm{Sp}}
\newcommand{\PSp}{\mathrm{PSp}}
     \newcommand{\SL}{\mathrm{SL}}
      \newcommand{\PSL}{\mathrm{PSL}}

    \renewcommand{\U}{\mathrm{U}}
     \renewcommand{\A}{\mathrm{A}}

\usepackage[new]{old-arrows}
    
    \newcommand{\tet}{2\mathrm{T}}
    \newcommand{\oct}{2\mathrm{O}}
    \newcommand{\ico}{2\mathrm{I}}

\newcommand{\bmsf}[1]{\bm{\mathsf{#1}}}

\renewcommand{\G}{\mathcal{G}}
\newcommand{\Glog}{\mathsf{G}}

\newcommand{\f}{\bmsf{f}}
\renewcommand{\F}{\bmsf{F}}

\newcommand{\irrGAP}{\bmsf{\chi}}

\newcommand{\logirr}{\bmsf{\lambda}}

\begin{document}
\title{Quantum Codes and Irreducible Products of Characters}
\thanks{These authors contributed equally to this work.}
\author{Eric Kubischta}
\email{erickub@umd.edu} 
\author{Ian Teixeira}
\email{igt@umd.edu}
\affiliation{Joint Center for Quantum Information and Computer Science,
NIST/University of Maryland, College Park, Maryland 20742 USA}
\begin{abstract}
In a recent paper, we defined twisted unitary $1$-groups and showed that they automatically induced error-detecting quantum codes. We also showed that twisted unitary $ 1 $-groups correspond to irreducible products of characters thereby reducing the problem of code-finding to a computation in the character theory of finite groups. Using a combination of GAP computations and results from the mathematics literature on irreducible products of characters, we identify many new non-trivial quantum codes with unusual transversal gates.
\end{abstract}

\maketitle

\section{Introduction}

In \cite{us3}, we introduced the notion of a \textit{twisted} unitary $t$-group. We showed that these led automatically to quantum codes of distance $d = t+1$ with precise transversal gate groups. We showed two examples: a twisted unitary $2$-group, the binary icosahedral group $\ico \subset \U(2)$, and a twisted unitary $1$-group, $\Sigma(360\phi) \subset \U(3)$. The codes produced by $\ico$ had distance $d = 3$, were realized by encoding a qubit into $n$ qubits, and could enact the group $\ico$ transversally. The codes produced by $\Sigma(360\phi)$ had distance $d=2$, were realized by encoding a qutrit into $n$ qutrits, and enacted $\Sigma(360\phi)$ transversally.

In this work we use the fact from \cite{us3} that twisted unitary $1$-groups are related to the irreducibility of a certain product of characters, or equivalently the irreducibility of a certain tensor product of representations. We use this simple criterion to find more examples of twisted unitary $1$-groups and thus many more examples of $d=2$ error-detecting codes. We also generalize the results from \cite{us3} by including codes that encode a qudit of size $K$ into $n$ qudits of size $q$. 

\section{Preliminaries}

\subsection{Twisted Unitary Groups}

Let $\G$ be a finite subgroup of $\U(q)$, the unitary group of degree $q$. Let $\F$ denote the fundamental representation of $\U(q)$ and let $\f$ denote the restricted representation of $\F$ to $\G$. Note that $\f$ is necessarily faithful.

Let $\logirr$ be an irreducible representation (irrep) of $\G$. Then we say that $\G$ is a $\logirr$-twisted unitary $t$-group if
\[
\frac{1}{|\G| } \sum\limits_{g \in \G} |\lambda(g)|^2  \qty( \f \otimes \f^*)^{\otimes t}(g)  = \int\limits_{\U(q)}  \qty(\F \otimes \F^*)^{\otimes t}(g) \, dg, \numberthis
\]
where the integral on the right is taken with respect to the unit-normalized Haar measure. Here $ \lambda $ denotes the character of $ \logirr $.

This definition is hard to work with but we showed in \cite{us3} that for $t=1$ there is a simple equivalent condition. Namely, $\G$ is a $\logirr$-twisted unitary $1$-group if and only if $|| \lambda f|| = 1$. In general we will use unbolded letters to denote the corresponding characters, so $ f $ and $ \lambda $ are the characters corresponding to $ \f $ and $ \logirr $. Also recall that for a character $ \chi $  we call $||\chi || := \expval{ \chi, \chi}$ the norm of the character. Although it is defined in terms of the standard inner product of characters $\expval{ \cdot, \cdot}$ in a way that we would usually think of as the norm squared, it is the convention in group theory to call this the norm. Recall that a product of characters corresponds to a tensor product of representations. For example, $\lambda f$ is the character of $\logirr \otimes \f$. Moreover, recall that an inner product of characters is $1$ if and only if the corresponding representation is irreducible. Thus $\G$ is a $\logirr$-twisted unitary $1$-group if and only if $\logirr \otimes \f$ is irreducible. 

Our goal in this paper is to find many more examples where $\logirr \otimes \f$, or equivalently $ \lambda f$, is irreducible. We described above a ``top down approach" - starting with $\U(q)$ and then selecting a finite subgroup $ \G $, and branching to identify the corresponding faithful irrep $\f$ and finally checking to see if there exist any $\logirr$ such that $\logirr \otimes \f$ is irreducible. However there is no classification of finite subgroups of $\U(q)$ in general and one would have to figure out the branching rules $\F$ to $\f$. So instead we will prefer a ``bottom-up approach." That is, we will start with a finite group $\G$ and allow $\f$ to be \textit{any} faithful irrep of $\G$. If $\f$ has degree $q$ then, since $\f$ was chosen to be faithful, $\G$ must be a subgroup of $\U(q)$. Then we compute $\logirr \otimes \f$ for each irrep $\logirr$ and see which are irreducible. 

Given a group $\G$, suppose for some $\f$ with degree $q$ and some $\logirr$ of degree $K$ we find that $\logirr \otimes \f$ is irreducible. Then $\G$ is a $\logirr$-twisted unitary $1$-group and so every $ \logirr $ subrepresentation of $ \f^{\otimes n} $ is a free $K$-dimensional quantum code with distance $d \geq 2$, i.e., these codes can detect an error on any single physical qudit. Moreover, these codes implement $\Glog := \logirr(\G)$ transversally (the image of $\G$ under $\logirr$). If $\logirr$ is faithful or almost faithful (meaning the kernel is small) then $\Glog$ will be a large set of transversal gates. The canonical way we find codes is by encoding a qudit of size $K$, via $\logirr$, into $n$-qudits of size $q$, via $\f^{\otimes n}$. There will be at least 1 good code if the multiplicity $\expval{\lambda, f^n} > 0$. We are most interested in the smallest $n$ this occurs for, and for convenience it is this smallest $ n $ that we will list in the tables below.

\subsection{Code Parameters and Transversality}

A general quantum code is denoted $ ((n,K,d))_q $ where $q^n $ is the dimension of the physical Hilbert space, $K$ is the dimension of the logical Hilbert space, and $d$ is the distance of the code. In other words, we are encoding a size $K$ logical qudit into $n$ size $q$ physical qudits, and the code can detect any error that affects at most $ d-1 $ physical qudits and correct any error that affects at most $\lfloor (d-1)/2 \rfloor$ physical qudits. If we are considering a stabilizer code, then we instead use the notation $ [[n,k,d]]_q $ where $K = q^k$. That is, we are encoding $k$ size $q$ logical qudits into $n$ size $q$ physical qudits.

Recall that for a $ ((n,K,d))_q $ quantum code a logical gate $ g \in \U(K)$ is called transversal if it can be implemented by a physical gate on at most one qubit from each code block, for example if it is enacted by a physical gate of the form $ U_1 \otimes \dots \otimes U_n $ with each $ U_i \in \U(q) $. Transversal gates are naturally fault tolerant because they act on physical qudits separately in such a way that a single initial error will always be propagated in a way that is correctable, even after arbitrarily many transversal gates are applied to the code. 

Since the holy grail of quantum error correction and fault tolerance is a universal fault tolerant quantum computer, it is desirable to have transversal gates which are as close to universal as possible (the no-go theorem of Eastin and Knill shows that transversal gates can never be universal \cite{EastinKnill}). One of the desirable properties of the Clifford group is that it is \textit{maximal} in $ \U(q^b) $, so a fault tolerant implementation of any single gate outside the Clifford group is all that is required to achieve a universal gate set. Another valuable property of the Clifford group is that (for $ q $ any prime power) it is a unitary $ 2 $-group, meaning that the gates are spread out in a way that well approximates all of $ \U(q^b) $. This feature likely lends itself to faster gate compilation times. 

An example of a quantum error-correcting code family that implements the Clifford Group transversally is $ b $ blocks of the Steane code, a $ [[7b,b,3]]_2 $ code whose transversal gate group is the $ b $ qubit Clifford group.

\subsection{Summary of Results} 
Naturally, one may be curious to find other $ ((n,K,d))_q $ codes whose transversal gate group $ \Glog $ is either maximal or a unitary $ 2 $-group in $ \U(K) $. In this paper we find many examples of such codes.

In section III we find a family of $ ((n_b,K_b,d))_{q_b} $ codes, indexed by $ b $, such that the size of the codespace, $ K_b \to \infty $ as $ b \to \infty $. Each code in this family implements the transversal gate group $ \Glog_b = \PSp(2b,3)$ which is a maximal unitary $ 2 $-group for $ \U(K_b) $. Thus this family is analogous to the Clifford-transversal Steane $ [[7b,b,3]]_2 $ code family.

In section IV we consider isolated examples of $ ((n,K,d))_q $ codes whose transversal gate group $ \Glog $ is a maximal unitary $ 2 $-group in $ \U(K) $. We can perform this task exhaustively using the results in \cite{tGroupsBannai}.

In section V, we explore other isolated examples of twisted unitary $1$-groups, most of which are found by a simple search with GAP.

In section VI, we exhaustively list twisting unitary $1$-groups appearing as subgroups of $\SU(2)$ and $\SU(3)$. These lead to free $ ((n,K,d))_q $ codes with small values of $ q $.

\section{$\PSp(2b,3)$ Code Family}

Let $ \bmsf{\eta_b} $ denote the degree $ |\bmsf{\eta_b}|= \tfrac{3^b-1}{2} $ Weil representation of $ \Sp(2b,3) $ and let $ \bmsf{\xi_b} $ denote the degree $ |\bmsf{\xi_b}|= \tfrac{3^b+1}{2} $ Weil representation of $ \Sp(2b,3) $. Then by \cite{tGroupsBannai} the $ \Sp(2b,3) $ subgroup of $ \U(\tfrac{q^n-1}{2}) $ is a unitary $ 2 $-group and, by the characteristic 3 subcase of prop 5.5 of \cite{MagaardTiep}, the $ \Sp(2b,3) $ subgroup of $ \U(\tfrac{q^n-1}{2}) $ is a $ \bmsf{\xi_b} $-twisted unitary $ 1 $-group. Similarly, the $ \Sp(2b,3) $ subgroup of $ \U(\tfrac{q^n+1}{2}) $ is a unitary $ 2 $-group and an $ \bmsf{\eta_b} $-twisted unitary $ 1 $-group.

So whenever $ \expval{\eta_b,\xi_b^n}>0 $ we get free $ d=2 $ quantum codes encoding a $ K_b=|\bmsf{\eta_b}|= \tfrac{3^b-1}{2} $ size codespace into $ n $ qudits each of size $ |\bmsf{\xi_b}|= \tfrac{3^b+1}{2} $. Similarly whenever $ \expval{\xi_b,\eta_b^n}>0 $ we get free $ d=2 $ quantum codes encoding a $ K_b=|\bmsf{\xi_b}|= \tfrac{3^b+1}{2} $ size codespace into $ n $ qudits each of size $ |\bmsf{\eta_b}|= \tfrac{3^b-1}{2} $.

Since $ |\bmsf{\xi_b}|=|\bmsf{\eta_b}|+1 $ then for each value of $ b $ one of the Weil modules is of even dimension ($ \bmsf{\eta_b} $ if $ b $ is even, $ \bmsf{\xi_b} $ if $ b $ is odd) and the other is of odd dimension. Weil representations of even dimension are always faithful but Weil representations of odd dimension are never faithful, instead having a kernel of size $ 2 $, and an image of $ \PSp(2b,3) $ rather than the original $ \Sp(2b,3) $. Since a faithful character can never occur in a power of a non-faithful character then for each value of $ b $ only one of the inner products $ \expval{\eta_b,\xi_b^n}, \expval{\xi_b,\eta_b^n} $ can be nonzero. And for sufficiently large $ n $ we are guaranteed that one of the inner products will be nonzero since every irrep occurs in a sufficiently large tensor power of any faithful irrep.

The smallest codes for the first few values of $ b $ are:

$ b=2 $, $ ((4,5,2))_4 $, $ \PSp(4,3) $-transversal

$ b=3 $, $ ((6,13,2))_{14} $, $ \PSp(6,3) $-transversal

This construction produces a family of codes with $ K_b \approx \tfrac{3^b}{2} $ and transversal gate group $ \Glog_b = \PSp(2b,3)$ a (maximal) unitary $ 2 $-group in $ \U(K_b) $. Thus we have produced a non-Clifford family of codes with $ K_b \to \infty $ as $ b \to \infty $ and transversal gate group $ \Glog_b $ always a (maximal) unitary $ 2 $-group in $ \U(K_b) $.

\subsection{Another Code Family}

Besides the Clifford groups for $ q $ a prime power, there are only two other infinite families of unitary $ 2 $-groups \cite{tGroupsBannai}. There is the family based on the Weil representations of $ \Sp(2b,3) $ described above, and there is a second family based on the Weil representations of $ \SU(b,2) $.

There are $ 3 $ Weil representations of $ \SU(b,2) $. The first representation, known as $ \bmsf{\zeta_{b,0}} $ is of degree $ |\bmsf{\zeta_{b,0}}|=\tfrac{2^b+2(-1)^b}{3}$ (which is always even). The other two representation $ \bmsf{\zeta_{b,1}}, \bmsf{\zeta_{b,2}} $ both have degree $ |\bmsf{\zeta_{b,1}}|=|\bmsf{\zeta_{b,2}}|=\tfrac{2^b-(-1)^b}{3} $ (which is exactly one dimension lower and so is always odd). By \cite{tGroupsBannai} the $ \SU(b,2) $ subgroup of $ \U(\tfrac{2^b-(-1)^b}{3})$ is a (maximal) unitary $ 2 $-group. We conjecture based on suggestive evidence given in \cite{MagaardTiep}, and confirmed in GAP for small $ b $, that the $ \SU(b,2) $ subgroup of $ \U(\tfrac{2^b-(-1)^b}{3})$ is a $ \bmsf{\zeta_{b,0}} $-twisted unitary $ 1 $-group. Thus every  $ \bmsf{\zeta_{b,1}} $ or $ \bmsf{\zeta_{b,2}} $ subrepresentation of $ \bmsf{\zeta_{b,0}}^{\otimes n} $ is a free $ d=2 $ code. When $ b $ is not divisible by $ 3 $ then $ \SU(b,2)= \PSU(b,2) $ is simple so all its Weil representations must be faithful, including $ \bmsf{\zeta_{b,0}} $, so we can always find sufficiently large $ n $ such that $ \expval{\zeta_{b,1},\zeta_{b,0}^n} >0 $, respectively $ \expval{\zeta_{b,2},\zeta_{b,0}^n} >0 $, and thus we can find the corresponding $ ((n,K_b,2))_{K_b+1}$ $ \SU(b,2) $ transversal codes, where $ K_b=|\bmsf{\zeta_{b,1}}|=|\bmsf{\zeta_{b,2}}|=\tfrac{2^b-(-1)^b}{3} $, and $ q_b=K_b+1= |\bmsf{\zeta_{b,0}}|=\tfrac{2^b+2(-1)^b}{3} $.

When $ b $ is divisible by $ 3 $ then the situation is significantly different: we have a central extension $ \SU(b,2) \cong 3.\PSU(b,2) $ and it is no longer the case that all representations of $\SU(b,2)$ are faithful. In particular $ \bmsf{\zeta_{b,0}} $ is no longer faithful, having a kernel of size $ 3 $, and no tensor powers of $ \bmsf{\zeta_{b,0}} $ contains $ \bmsf{\zeta_{b,1}} $ or $ \bmsf{\zeta_{b,2}} $. Thus we are not able to obtain any codes this way that have a unitary $ 2 $-group as the transversal gate group.

Note however that we can still get free codes in the $ b $ divisible by $ 3 $ case by encoding  $ \bmsf{\zeta_{b,0}} $ into powers of $ \bmsf{\zeta_{b,1}} $ or $ \bmsf{\zeta_{b,2}} $. Although now the codespace is one dimension larger and the $ \SU(b,2) $ subgroup is no longer a $ 2 $-group or a maximal subgroup.

\section{Finding More Codes when $ \Glog $ is a unitary $ 2 $-group }

In this section we consider exceptional unitary $ 2 $-groups $ \Glog $ that are outside of the infinite families. All codes described in this section have a transversal gate group which is maximal in $ \U(K) $ and is a unitary $ 2 $-group (or better).

Given $ \Glog $ a (maximal) unitary $ 2 $-group in $ \U(K) $ we attempt to find $ \Glog $ transversal $ ((n,K,2))_q $ codes by searching for a $ \logirr $-twisted unitary $ 1 $-group $ \G $ with $ \logirr(\G)=\Glog $.

This approach is remarkably successful and we find appropriate twisted $ 1 $-groups for 14 of the 19 exceptional unitary $ 2 $-groups. The total of 19 comes from the 15 in table I of \cite{tGroupsBannai} as well as $ 2.\A_5 \subset \SU(2)$, $3.\A_6 \subset \SU(3)$, $\SL(3,2) \subset \SU(3)$, $ 2.\A_7 \subset \SU(4)  $ (while noting that $ \Sp(4,3) \subset \SU(4) $ is not exceptional because it is from a Weil representation). The five remaining $ 2 $-groups we do not find codes for this way are $ 6.\A_7 \subset \SU(6)$, $ 6.\PSL(3,4).2_1 \subset \SU(6)$, $ 6_1.\PSU(4,3).2_2 \subset \SU(6)$, $ Sz(8).3 \subset \SU(14)$, $J_4 \subset \SU(1333)$.

\begin{table}[htp] 
 \small  
    \begin{tabular} {lllll} \toprule  
    $ K $ & $\Glog =\logirr(\G)$ & $ N=q^n $ & $ \G $ & $ t $   \\ 
    \toprule 
   2 & $  2.\A_5 $ & $2^7$ & $ 2.\A_5 $ & 5 \\ \addlinespace
   \midrule
  3 & $ 3.\A_6 $ & $3^5$ & $ 3.\A_6 $ & 3  \\ 
    3 & $ 3.\A_6 $ & $3^7$ & $ 3.\A_6 $ & 3  \\ 
   3 & $ \SL(3,2) $ & $7^4$ & $ 2^3 \rtimes \SL(3,2) $ & 2 \\
   \midrule
   4 & $ 2.\A_7 $ & $6^9 $ & $ 6.\A_7 $ & 3 \\
   \midrule
      5 & $ \PSp(4,3) $ & $4^4 $ & $ \Sp(4,3) $ & 2 \\
       5 & $ \PSp(4,3) $ & $4^6 $ & $ \Sp(4,3) $ & 2  \\
      5 & $ \SU(4,2) $ & $6^6 $ & $ \SU(4,2) $ & 2  \\
\midrule
8 & $ 4_1.\PSL(3,4).2_3 $ & $8^5 $ & $ 4_1.\PSL(3,4).2_3 $ & 2 \\
\midrule
10 & $ 2.M_{12}.2 $  & $12^5 $ & $ 2.M_{12}.2 $ & 2 \\
10 & $ 2.M_{22}.2 $ & $56^3 $ & $ 2.M_{22}.2 $ & 2 \\
\midrule
11 & $ \SU(5,2) $ & $10^5 $ & $ \SU(5,2) $ & 2 \\
\midrule
12 & $ 6.Suz $ & $11088^5 $ & $ 6.Suz $ & 2 \\
\midrule
13 & $ \PSp(6,3) $ & $14^6 $ & $ \Sp(6,3) $ & 2 \\
\midrule
18 & $ 3.J_{3} $ & $18^5 $ & $ 3.J_{3} $ & 3 \\
\midrule
26 & $ ^2F_4(2)' $ & $27^5 $ & $ ^2F_4(2)' $ & 2 \\
\midrule
28 & $ 2.Ru $ & $1248^3 $ & $ 2.Ru $ & 2 \\
\midrule
43 & $ \SU(7,2) $ & $42^9 $ & $ \SU(7,2) $ & 2 \\
\midrule
45 & $ M_{23} $ & $22^5 $ & $ M_{23} $ & 2 \\
45 & $ M_{24} $ & $23^6 $ & $ M_{24} $ & 2 \\
\midrule
342 & $ 3.ON $ & $ 495^4 $ & $ 3.ON $ & 2 \\
  \addlinespace \bottomrule
    \end{tabular}
    \caption{Table of free codes with transversal unitary $ t $-groups, $ t \geq 2 $}
    \label{tab:2designcodes}
\end{table}

\cref{tab:2designcodes} displays the following data: the size $ K $ of the codespace, the transversal gate group $ \Glog=\logirr(\G) \subset \U(K) $, the size $ N=q^n $ of the physical space for the smallest free code, the $ \logirr $-twisted unitary $ 1 $-group $ \G \subset \U(q) $, the value $ t\geq 2 $ for which $ \Glog $ is a unitary $ t $-group in $ \U(K) $.

Note that the rows with $ \G=\Sp(4,3), \Sp(6,3) $ correspond to the codes from Weil representations of $ \Sp(2b,3) $, already discussed in the first section. Similarly, the rows with $ \G= \SU(4,2), \SU(5,2), \SU(7,2) $ correspond to the codes from Weil representations of $ \SU(b,2) $ already discussed.


In all cases, $ q=|\f| $ is the degree of the fundamental irrep and the size of the codespace $ K=|\logirr| $ is the degree of the logical irrep. The value of $ n $ given is just the least $ n $ such that $ \expval{\lambda,f^n}>0$. There will be infinitely many higher tensor powers of $ \f $ containing $ \logirr $ subreps, and all such $ \logirr $ subreps will correspond to $ d>1 $, error-detecting codes. For more information about these moduli spaces of free quantum codes see \cite{us3}.

\section{Finding Codes when $ \Glog $ is not a unitary $ 2 $-group }

In this section we use $ \logirr $-twisted unitary $ 1 $-groups to find other free quantum codes with large transversal gate group $ \Glog $.

\begin{table}[htp] 
 \small  
    \begin{tabular} {llll} \toprule  
    $ K $ & $\Glog=\logirr(\G)$ & $ N=q^n $ & $ \G$ \\ 
   \midrule
   3 & $ \A_5 $ & $2^6$ & $ 2.\A_5 $  \\
   \midrule
   4 & $ \SL(2,9)\cong 2.\A_6 $ & $15^3 $ & $ 2^5 \rtimes \A_6 $   \\
   4 & $ \SL(2,9).2\cong 2.\S_6 $ & $15^3 $ & $ 2^5 \rtimes \S_6 $  \\
   \midrule
   5 & $ \A_6 $ & $3^6 $ & $ 3.\A_6 $ \\
    \midrule
   6 & $ 3.\A_7 $ & $6^4 $ & $ 6.\A_7 $ \\
   6 & $ \A_7 $ & $6^6 $ & $ 6.\A_7 $  \\
   6 & $ 2.J_2 $ & $ 6^9 $ & $ 2.J_2 $  \\
    \midrule
   7 & $ \A_8\cong \SL(4,2) $ & $15^3 $ & $ 2^4 \rtimes \SL(4,2) $  \\
\midrule
8 & $ \A_9 $ & $21^4 $ & $ \A_9 $   \\
\midrule
14 & $ G_2(3) $ & $27^6 $ & $ 3.G_2(3) $   \\
\midrule
22 & $ McL $ & $126^3 $ & $ 3.McL $  \\
22 & $ 2.HS.2 $ & $56^4 $ & $ 2.HS.2 $  \\ 
\midrule
23 & $ Co2 $ & $9625^3 $ & $ Co2 $  \\
23 & $ Co3 $ & $896^4 $ & $ Co3 $  \\
\midrule
24 & $ 2.Co1 $ & $9152000^3 $ & $ 2.Co1 $  \\
\midrule
27 & $ 3.G_2(3) $ & $27^4 $ & $ 3.G_2(3) $  \\
\midrule
78 & $ Fi22 $ & $ 352^6 $ & $ 2.Fi22 $ \\
\midrule
248 & Thompson & $ 27000^4 $ & Thompson  \\
\midrule
4371 & Baby Monster & $ 53936390144^4 $ & Baby Monster \\
\midrule
8671 & $ Fi24 $ & $ 1603525^3 $ & $ Fi24 $  \\
\midrule
196,883 & Monster Group & $ 8980616927734375^3 $ & Monster Group  \\ 
  \addlinespace \bottomrule
    \end{tabular}
    \caption{Table of small free codes for other notable groups $ \G \subset \SU(K) $ (not 2-groups)}
    \label{tab:corr}
\end{table}

Whereas in the previous sections we essentially went through the classification of unitary $ t $-groups $ \Glog $ given in \cite{tGroupsBannai} and found that surprisingly many of them arise as $ \logirr(\G) $ for $ \G $ some $ \logirr $-twisted unitary $ 1 $-group, now we do not have a list of desired $ \Glog $ to guide us. Instead we just list a number of large $ \G $ which happen to be twisted $ 1 $-groups, either by proof given in the literature on irreducible products of characters, or by direct check in GAP. We list our results in \cref{tab:corr}.

As usual, if $ \G $ is a $ \logirr $-twisted $ 1 $-group with $ \logirr $ faithful then $ \G= \logirr(\G)=\Glog $. On the other hand if $ \logirr $ is not faithful then the group of transversal gates $ \Glog $ is just $ \G $ mod some normal subgroup.

It is worth remembering that each row in this table is really an infinite family of $ d>1 $ quantum codes with larger and larger moduli spaces of codes occurring for infinitely tensor powers of $ \f $ beyond the minimal power $ n $ such $ \expval{\lambda,f^n}>0 $, which is the value of $ n $ given in the table. 

Some of the twisted unitary $ 1 $-groups in this table are from \cite{NavarroTiep2021,MagaardTiep}. $ \A_{m^2} $ for all $ m \geq 3 $ has an irreducible product of characters \cite{Zisser_1993}, the codes which can be produced this way show up for example in the row with $ \G=\A_9 $.

There are also conjectured code families for $ \Sp(2b,5) $ and $ \Sp(2b,2) $, again using the corresponding Weil representations, see Theorem D of \cite{NavarroTiep2021}.

\section{Free codes using physical qubits and qutrits}

In this last section we exhaustively list the twisted unitary $1$-groups that are subgroups of either $\SU(2)$ or $\SU(3)$; these lead to codes involving just physical qubits and qutrits. These do not have especially large transversal gate groups (or if they do they were already covered in a previous section or in \cite{us3}) but they are of practical interest because they are made with physical qudits of small size.

For ease, if $\logirr \otimes \f$ is irreducible we will say that the pair $(\f, \logirr)$ has the \textit{automatic protection property} (or simply the APP).

\subsection{APP in $\SU(2)$}
Let us start with the qubit case, $q = 2$. We find that only two subgroups of $\SU(2)$ permit APP pairs: $\oct$, the binary octahedral group and $\ico$, the binary icosahedral group.

\subsubsection{$\G = \ico$} 
Let us start with $\ico$ since that group was also studied in \cite{us3}. This unitary $5$-group is also known as $\SL(2,5)=\Sp(2,5)$ or $2.\A_5$ and we can access it in GAP as \texttt{PerfectGroup(120)}.

In \cite{us3} we already saw that $\f = \irrGAP_2$ and $\logirr = \irrGAP_3$ was an APP pair (and in fact yielded a twisted unitary $2$-group). But we could have just as easily used the reverse: $\f = \irrGAP_3$ and $\logirr = \irrGAP_2$ is also an APP pair. Both of these cases yield a $((7,2,3))_2$ quantum code with transversal $\ico$. The idea is that $\irrGAP_2$ and $\irrGAP_3$ (both of degree 2) are Galois conjugates (and related by an outer automorphism) so either is a valid choice as a fundamental irrep for $\SU(2)$ (and there really is no effective difference between these cases). We call these APP pairs \textit{equivalent}.

On the other hand, there is another APP pair up to equivalency: $\f = \irrGAP_2$ and $\logirr=\irrGAP_4$ (or $\f = \irrGAP_3$ and $\logirr = \irrGAP_5$). In this case, $\f$ has degree $2$ and $\logirr$ has degree $3$. These correspond to codes where we encode a logical qutrit into $n$-qubits. The smallest occurs when $n = 6$, i.e., 
a $((6,3,2))_2$ code that encodes a qutrit into 6 qubits and can detect any single error. Note that this is entirely different from the previous scenario: we see that the degree of $\f$ corresponds to $q$ and the degree of $\logirr$ corresponds to $K$. Also new is that $\logirr$ is not a faithful irrep, the image $\Glog = \logirr(\G)$ is isomorphic to $\A_5$, the alternating group on $5$-letters, and this is a well known subgroup of $ \SU(3) $. Thus this $((6,3,2))_2$ code can implement $\A_5$ transversally.

\subsubsection{$\G = \oct$}
Now let us move onto $\oct$, also known as $2.\mathrm{S}_4$. We can call this unitary $3$-group in GAP via \texttt{SmallGroup(48,28)}. 

There is one APP pair up to equivalency: $\f = \irrGAP_4$ and $\logirr = \irrGAP_3$ (or equivalently $\f = \irrGAP_5$ and $\logirr = \irrGAP_3$). Because both irreps have degree 2 this is the case that we encode a qubit into $n$-qubits and the smallest code occurs in $n = 4$, i.e., we have a free $((4,2,2))_2$ code. However, $\irrGAP_3$ is extremely non-faithful and so $\Glog$ is only $\mathrm{S}_3$, the symmetric group on 3-letters.

\subsection{APP in $\SU(3)$}

Now we consider the case that $\G$ is a subgroup of $\SU(3)$. The finite subgroups of $\SU(3)$ are listed in \cite{SUtree} and we follow the notation therein. There are only 5 subgroups here that admit APP pairs: 
\[
    \Sigma(72\phi), \quad \Delta(6 \cdot 6^2), \quad \Delta(6\cdot 9^2), \quad \Sigma(216\phi), \quad \Sigma(360\phi).
\]

\subsubsection{$\G=\Sigma(72\phi)$}
This unitary $2$-group can be called in GAP via \texttt{SmallGroup(216,88)}. This group admits one APP pair (up to equivalence) given by $\f = \irrGAP_6$ and $\logirr = \irrGAP_5$. The smallest code here is a $((3,2,2))_3$ code which encodes a qubit into 3 qutrits and can detect any single error. This code implements the single qubit Pauli group transversally.

\subsubsection{$\G=\Delta(6 \cdot 6^2)$} 
This group can be called in GAP via \texttt{SmallGroup(216,95)}. This group admits one APP pair (up to equivalence): $\f = \irrGAP_{13}$ and $\logirr= \irrGAP_4$. The smallest code here is a $((6,2,2))_3$ code which encodes a qubit into 6 qutrits and detects any single error while implementing the group $C_3$ transversally, the cyclic group on 3-letters.

\subsubsection{$\G=\Delta(6 \cdot 9^2)$} 
This group can be called in GAP via \texttt{SmallGroup(486, 61)}. This group admits one APP pair (up to equivalence): $\f = \irrGAP_{11}$ and $\logirr = \irrGAP_4$. The smallest code here is a $((9,2,2))_3$ code which encodes a qubit into 9 qutrits and detects any single error while implementing the group $C_3$ transversally, the cyclic group on 3-letters.

\subsubsection{$\G=\Sigma(216\phi)$}
This unitary $2$-group is the qutrit Clifford group and can be called as \texttt{SmallGroup(648, 532)}. This group admits four APP pairs (up to equivalence). The 1st pair is $\f = \irrGAP_8$, $\logirr = \irrGAP_6$ and the smallest code here is $((6,2,2))_3$ that encodes a qubit into 6 qutrits and can implement the qubit Pauli group transversally. The 2nd pair is $\f = \irrGAP_8$, $\logirr = \irrGAP_5$ and the smallest code here is $((3,2,2))_3$ that encodes a qubit into 3 qutrits and can implement the qubit Pauli group transversally. The 3rd pair is $\f = \irrGAP_8$, $\logirr = \irrGAP_4$ and the smallest code here is $((6,2,2))_3$ that encodes a qubit into 6 qutrits and can implement the group $\tet = \SL(2,3)$ transversally (which is a subgroup of the qubit Clifford group). The 4th APP here is $\f = \irrGAP_8$, $\logirr = \irrGAP_7$ and the smallest code here is $((6,3,2))_3$ that encodes a qutrit into 6 qutrits and can implement the group $A_4$ transversally (the alternating group on $4$ letters).

\subsubsection{$\G=\Sigma(360\phi)$} 
We studied this unitary $3$-group in \cite{us3} and it can be called in GAP via \texttt{PerfectGroup(1080)}. This group admits three APP pairs (up to equivalence). We already found 2 of these pairs in \cite{us3}: the 1st pair is $\f = \irrGAP_2$, $\logirr = \irrGAP_3$ and the smallest code here is $((7,3,2))_3$ that encodes a qutrit into 7 qutrits and can implement $\Sigma(360\phi)$ transversally. The 2nd pair is $\f = \irrGAP_2$, $\logirr = \irrGAP_4$ and the smallest code here is $((5,3,2))_3$ that encodes a qutrit into 5 qutrits and can implement $\Sigma(360\phi)$ transversally.

The 3rd pair is new. It is given (up to equivalence) by $\f = \irrGAP_2$, $\logirr = \irrGAP_6$ and the smallest code here is $((6,5,2))_3$ that encodes a qudit of size $5$ into 6 qutrits and can implement the group $A_6 \subset \SU(5)$ transversally (the alternating group on $6$ letters). 

\section{Conclusion}
Using a combination of GAP and the existing literature on irreducible products of characters, we have found many $ \logirr $-twisted unitary $ 1 $-groups. By the main result of \cite{us3} these correspond to free quantum codes. Indeed, using this approach we find $ ((n,K,2))_q $ codes with transversal gate group $ \Glog $ for nearly all unitary $ 2 $-groups $ \Glog \subset \U(K) $. We also find many other codes with large transversal gate group $ \Glog $. And where there exists a full classification of the finite subgroups of $ \U(q) $ for $ q=2,3 $ we find all possible irrep pairs that lead to free codes. This plethora of novel codes showcases the strength of the results obtained in \cite{us3}.

\section{Acknowledgments} 
This research was supported in part by the MathQuantum RTG through the NSF RTG grant DMS-2231533.

\bibliography{biblio}

\end{document}